\documentclass[%
reprint,
 amsmath,amssymb,
 aps,
 onecolumn,
aps,
]{revtex4-2}

\usepackage[utf8]{inputenc}
\usepackage{amsmath}
\usepackage{graphicx}
\usepackage{appendix}
\usepackage{color}
\begin{document}
\title{R\'enyi Entropy of Zeta-Urns}
\author{Piotr Bialas}\email{piotr.bialas@uj.edu.pl}
\affiliation{Institute of Applied Computer Science, Jagiellonian University,\\ ul. Lojasiewicza 11, 30-348 Krak\'ow, Poland}
\author{Zdzislaw Burda}\email{zdzislaw.burda@agh.edu.pl} 
\affiliation{AGH University of Krakow, Faculty of Physics and Applied Computer Science, \\
al. Mickiewicza 30, 30-059 Krak\'ow, Poland}
\author{Desmond A. Johnston}\email{D.A.Johnston@hw.ac.uk}
\affiliation{School of Mathematical and Computer Sciences, Heriot-Watt University,\\ Riccarton, Edinburgh EH14 4AS, UK }

\date{\today}


\begin{abstract}
We calculate analytically the R\'enyi entropy for
the zeta-urn model with a Gibbs measure definition of the micro-state probabilities. This allows us to obtain the singularities in the R\'enyi entropy from those of the thermodynamic potential, which is directly related to the free energy density of the model. We enumerate the various possible behaviours of the R\'enyi entropy and its singularities, which depend on both the value of the power law in the zeta-urn and the order of the R\'enyi entropy under consideration.
\end{abstract}

\maketitle

\section{Introduction}
Diversity, and how to measure it, has been a subject of fundamental interest in mathematical biology and ecology for many years \cite{pt,jbws,ld,chj,lc,lj,ht,hu,bhrse}. There have been interesting contributions from numerous authors that make use of ideas from statistical mechanics and thermodynamics, specifically those related to various notions of entropy. A prototypical problem is to quantify the diversity of an ecosystem whose organisms may be divided into $N$ distinct species, where the $\sigma$th species has a relative abundance of $p_{\sigma}$, so
\begin{equation}
p_1 + p_2 + \cdots p_N = 1 \, .
\end{equation}
From a statistical mechanical perspective $p_{\sigma}$ is the probability of having a micro-state $\sigma$ in some ensemble.
Given the $p_{\sigma}$ there are a multitude of entropy-like measures of diversity that one might consider (and which have already been proposed), a small selection being:
\begin{eqnarray}
\label{diversities}
\text{Species richness} &:&\sum_{\sigma} p_{\sigma}^0 \nonumber  \\
\text{Shannon entropy \cite{cs}} &:&-\sum_{\sigma} p_{\sigma} \log p_{\sigma}\nonumber \\
{}\\
\text{Gini-Simpson Index \cite{si}} &:&1 - \sum_{\sigma} p_{\sigma}^2\nonumber \\
\text{Tsallis  Entropy (of order $\lambda$) \cite{te}} &:&\frac{1 - \sum_{\sigma} p_{\sigma}^{\lambda}}{\lambda-1}\nonumber \\
\text{R\'enyi Entropy (of order $\lambda$),  } H_{\lambda} \text{ \cite{r}} &:&  \frac{1}{1 -\lambda} \log \sum_{\sigma} p_{\sigma}^{\lambda}\nonumber 
\end{eqnarray}
The parameter $\lambda$ is a non-negative real number.
In the limit $\lambda \rightarrow 1$, the R\'enyi entropy, 
$H_{\lambda}$, \cite{r} reproduces the Shannon entropy, 
$H_1 = -\sum_\sigma p_\sigma \log p_\sigma$, \cite{cs} 
and in the limit $\lambda \rightarrow 0$, it gives
the logarithm of the species richness (i.e. logarithm of the number of micro-states): 
$H_0 = \log \sum_{\sigma} p_{\sigma}^0  = \log \sum_\sigma 1$. 

We will focus on the R\'enyi entropy here. Its exponential, the diversity or Hill number \cite{hn} of order $\lambda$, is denoted by $D_{\lambda} = \exp( H_{\lambda}) $
\begin{equation}
D_{\lambda} (\bar{p}) = \left( \sum_{\sigma} p_{\sigma}^{\lambda} \right)^{\frac{1}{1-\lambda}}
\end{equation}
where we have defined an abundance vector $\bar{p} = ( p_1, p_2, p_3, \ldots, p_N )$.  The Hill number
is perhaps a more suitable choice than the entropy itself in an ecological setting since the resulting Hill numbers generally have a direct interpretation in terms of familiar quantities. For instance,  $D_0(\bar{p})$ will be the number of distinct
species and
\[
D_2(\bar{p}) = \frac{1}{\sum_{\sigma} p_{\sigma}^2}
\]
is the inverse participation ratio.
Also, in the uniform case $D_{\lambda}(1/N,1/N,\ldots,1/N)=N$, \,  $\forall \lambda$, giving the number of species.  In essence, the Hill numbers and their generalizations are providing an ``effective number of species'' for an ecosystem with some input from our prejudices on the importance of rare species determined by the parameter $\lambda$. 
The parameter $\lambda$ can be thought of as tuning the sensitivity of the diversity measure $D_{\lambda} (\bar{p})$ to the occurrence of rare species. Since the summands are given by $p_{\sigma}^{\lambda}$, rare species (smaller $p_{\sigma}$) will be weighted less strongly as $\lambda$ is increased. The highest sensitivity to rare species is therefore given by $\lambda=0$.

It is possible to further refine (complicate!)  such models by introducing a measure of the similarity between $Z_{\sigma \nu}$ between species $\sigma$ and $\nu$, with $0 \le Z_{\sigma \nu} \le 1$, where $Z_{\sigma \nu}=0$ is total dissimilarity and $Z_{\sigma \nu}=1$ is total similarity \cite{lc}. In this case the Hill numbers would be modified to 
\begin{equation}
D_{\lambda}^Z (\bar{p}) = \left( \sum_{\sigma} p_{\sigma} \left( ( Z \cdot \bar{p})_{\sigma} \right)^{\lambda-1} \right)^{\frac{1}{1-\lambda}}
\end{equation}
 where
\begin{equation}
( Z \cdot \bar{p})_{\sigma} = \sum_{\nu} Z_{\sigma \nu} p_{\nu}
\end{equation}
We shall consider only the case  $Z=I$ here.

It has been observed   \cite{b,mw} that if the $p_{\sigma}$ are given by a Gibbs measure
\begin{equation}
p_{\sigma} = \frac{\exp ( - \beta E_{\sigma} )}{Z(\beta)}
\end{equation}
with the partition function defined by $Z(\beta)= \sum_{\sigma} \exp (- \beta E_{\sigma})$, then the R\'enyi entropy $H_\lambda$ is related to the logarithm of the ratio
of partition functions  
\begin{equation}
    H_\lambda = \frac{1}{1-\lambda} \log \frac{Z(\lambda \beta)}{Z^\lambda(\beta)} \, .
    \label{HZ}
\end{equation}
This may also be written as a difference of free energies, $F(\beta) = - (1/\beta) \log Z (\beta)$,
\begin{equation}
H_{\lambda} =  \lambda \beta^2 \left[\frac{F(\lambda \beta ) - F (\beta)}{\lambda \beta - \beta} \right]
\label{HF}
\end{equation}
where the expression is the square brackets
may be regarded as a $q$-derivative of $F$   defined by
\[
\left(\frac{dF(x)}{dx}\right)_q = \frac{F(qx)- F(x)}{qx -x} \, .
\label{qdiff}
\]
(with $\lambda$ playing the role of $q$).
We recover the usual relation between the Shannon entropy and the free energy in the limit of $\lambda \to 1$. This relation is also the basis of using the R\'enyi entropy and the replica trick in conformal field theory \cite{cc} and numerical \cite{fglt} calculations to evaluate the entanglement entropy of various statistical mechanical systems.

\section{The model}
It is tempting to use simple explicitly solvable statistical mechanical models to investigate the properties of diversity measures such as the R\'enyi  entropy and, indeed, this has already been done in \cite{maac} for the class of models, zeta-urns, which we address here. However, our aims and also our notion of a ``species''/micro-state, are rather different from those of \cite{maac} as we highlight below.

Zeta-urn models describe weighted partitions of $S$ balls (particles) between $N$ boxes, such that each box $i$ contains at least one particle, $s_i\ge 1$, and  $S=s_1+s_2+\ldots + s_N$.
In our case micro-states $\sigma$ in this model 
correspond to the particle distributions in the boxes 
$\sigma = (s_1,s_2,\ldots,s_N)$.
The number of states for $S$ particles in $N$ boxes is 
$\mathcal{N} = \binom{S-1}{N-1}$. The abundance vector 
$(p_1,....,p_{\mathcal{N}})$ 
is thus not the same as in \cite{maac}, where the authors considered the abundance vector 
$(p_1,....,p_S)$  in the set $\{1,2,...,S\}$ in an ensemble in which the number of boxes $N$ was allowed to fluctuate. The  geometrical picture behind this choice in \cite{maac} is of breaking a bar of length $S$ into $N$ segments of sizes  $(s_1,s_2,\ldots,s_N)$ and maximizing the diversity (by some measure) of these, which was then applied to the general problem of partitioning a set of $S$ elements into $N$ components with a power-law distribution for the probabilities of the component sizes.
In \cite{maac} it was found using a phenomenological calculation based on cluster size estimation in spin models \cite{adal} that $D_0$ was maximized for  $\beta=2$, which is the value given by  Zipf's law \cite{gkz,mejn}. This, and the predicted scaling with $S$, agreed well with numerical simulations.

Here we would like to use (\ref{HZ},\ref{HF}) to investigate the singular behaviour 
of the R\'enyi entropies in a zeta-urn model. To this end, the energy of the system in the  state $\sigma$ is taken  to be logarithmic in the number of particles in each box
\begin{equation}
    E_\sigma = \sum_{i=1}^N \log s_i \, . 
    \label{Esig}
\end{equation}
The corresponding partition function \cite{bbj,dgc} 
\begin{equation}
    Z_{S,N}(\beta) = \sum_\sigma e^{-\beta E_\sigma}
    \label{Zsig}
\end{equation}
may be rewritten as 
\begin{equation}
    Z_{S,N}(\beta) = \sum_{(s_1,\ldots,s_N)} 
    w(s_1) \ldots w(s_N) \delta_{S- (s_1+\ldots +s_N)}
    \label{ZSN}
\end{equation}
with 
\begin{equation}
    w(s) = s^{-\beta} 
    \label{plw}
\end{equation}
for $s=1,2,3,\ldots$.  The parameter $\beta$ in the power law for the weights can thus be considered as the inverse 
temperature: $\beta=1/T$. Despite its simplicity, this model
occurs in many problems of statistical physics, including zero range processes \cite{zrp,e,gl,g} (as a non-equilibrium steady state), mass transport \cite{mez,emz,emz2},
random trees \cite{bb,j}, lattice models of quantum gravity \cite{bbj2,bb2,bbw}, emergence of the longest interval 
in tied down renewal processes \cite{g1,g2}, 
wealth condensation \cite{bjjknpz} and diversity 
of Zipf's population \cite{maac}. The system
described by the model has a phase transition which 
is associated with a real-space condensation \cite{bbj}.  

We will study the R\'enyi entropy for this model with a given micro-state being a particle
distribution in the boxes $\sigma  = (s_1, s_2, . . . , s_N )$ as described above.
The R\'enyi entropy is defined as in Eq. (\ref{diversities})
\begin{equation*}
    H_\lambda = \frac{1}{1-\lambda} \log \sum_{\sigma} p^\lambda_\sigma   \, ,
    \label{Hl}
\end{equation*}
where $p_\sigma$ is the probability of the $\sigma$ state:
\begin{equation}
    p_\sigma = \frac{1}{Z(\beta)} e^{-\beta E_\sigma} .
    \label{ps}
\end{equation}
$Z$ in the last equation is an abbreviation for $Z_{S,N}(\beta)$.  With the Gibbs measure definition of the micro-states employed here the free-energy difference/R\'enyi entropy relations of (\ref{HZ},\ref{HF}) apply. This in turn allows us to relate the singular behavior of the R\'enyi entropy (density) to that of the  free energy (density).

Our aim is to calculate $H_\lambda$ explicitly in the thermodynamic limit
\begin{equation}
   S\rightarrow \infty , \  \frac{N}{S} \rightarrow r  \, , 
   \label{dlimit}
\end{equation}
where $r \in (0,1)$, and then use this to obtain the singular behaviour, if it exists. The parameter $r$ is a free parameter which is equal to the inverse particle density 
(i.e. the average number of particles per box).
In the limit (\ref{dlimit}), the free energy $F(\beta,r,S)$ is
an extensive quantity which means that it grows 
linearly with the system size $F(\beta,r,S) = S f(\beta,r) + o(S)$ as $S$ goes to infinity. We are interested 
in the coefficient $f(\beta,r)$ of the leading term,
which can be interpreted as the free energy per particle. 
Only if this coefficient is zero need the next-to-leading 
terms denoted by $o(S)$ be considered.  In general, we will use
the convention that extensive quantities will be denoted
by capital letters, while the corresponding densities
by small-case letter. In particular, we will denote the
R\'enyi entropy per particle (or the R\'enyi entropy density) 
by $h_\lambda$.

Let us introduce a thermodynamic potential
\begin{equation}
 \phi(\beta,r) = \lim \frac{1}{S} \log Z_{S,N} 
\label{phi}
\end{equation}
where ``lim'' in this equation means the  thermodynamic limit as defined in (\ref{dlimit}). The function $\phi(\beta,r)$
gives the rate of exponential growth of the partition 
function with $S$ in the thermodynamic limit (\ref{dlimit}),
and it is directly related to the free energy density: 
$\phi(\beta,r) = - \beta f(\beta,r)$. Dividing both sides
of (\ref{HZ}) by $S$ and taking the limit (\ref{dlimit})
we find a direct relationship between the R\'enyi entropy density and the thermodynamic potential $\phi$
\begin{equation}
    h_\lambda(\beta,r) = 
    \frac{\phi(\lambda \beta,r) - \lambda\phi(\beta,r)}{1-\lambda} .
    \label{hphi}
\end{equation}
For $\lambda \rightarrow \infty$ the last equation
reduces to $h_\infty(\beta,r)=\phi(\beta,r)$ while for
$\lambda=0$ to $h_0(\beta,r)=\phi(0,r)$. Clearly, $h_0(\beta,r)$ is independent of $\beta$. It can be
easily determined by enumeration of states which gives
\begin{equation}
    Z_{S,N}(0) = \sum_\sigma 1 = \binom{S-1}{N-1} = \mathcal{N} .
\end{equation}
Substituting this into (\ref{phi}) we find 
in the thermodynamic limit (\ref{dlimit}) 
\begin{equation}
h_0(\beta,r) = \phi(0,r) =  -r \log r - (1-r) \log(1-r) . 
\label{h0}
\end{equation}
For the Shannon entropy (density) limit $\lambda \rightarrow 1$
\begin{equation}
    h_1(\beta,r) = \phi(\beta,r)-\partial_\beta \phi(\beta,r) 
    \label{h1}
\end{equation}
as can be seen by applying L'Hôpital's rule to
(\ref{hphi}).

The thermodynamic potential $\phi(\beta,r)$ can be found
analytically using the saddle point method. 
The details can be found in \cite{bbj,dgc,bbj2}  or  
in a more recent paper where  results on the phase structure 
of the zeta-urn model are updated and collected in one place \cite{balls2}.
Here we quote the result which is expressed in terms of a generating function 
\begin{equation}
   K_\beta(\alpha) = \log \sum_{k=1}^\infty w(k) e^{-\alpha k} = \log {\rm Li}_\beta(e^{-\alpha}) 
   \label{kalpha}
\end{equation}
where ${\rm Li}_\beta(z)$ is the polylogarithm:
\begin{equation}
    {\rm Li}_\beta(z) = \sum_{k=1}^\infty \frac{z^k}{k^\beta} .
\end{equation}
The middle expression in \eqref{kalpha} is
a definition of the generating function, while the
last expression is just its explicit form
for the power-law weights \eqref{plw}.

For $\alpha=0$
\begin{equation}
   K_\beta(0) = \log \zeta(\beta) . 
\end{equation}
Two cases can be distinguished. For $\beta \le 2$, $\phi(\beta,r)$ can be expressed as
parametric equations, where both $\phi$
and $r$ are parameterized by $\alpha \in (0,\infty)$:
\begin{equation}
    r = -\frac{1}{K_\beta'(\alpha)}
    \label{pr1}
\end{equation}
and
\begin{equation}
    \phi(\beta,r) = \alpha - \frac{K_\beta(\alpha)}{K_\beta'(\alpha)} .
    \label{pr2} 
\end{equation}
These equations are valid for all values from the range $r \in (0,1)$,
which is an image of the range $\alpha \in (0,\infty)$ 
of the mapping (\ref{pr1}).

For $\beta>2$, the image of the range $\alpha \in (0,\infty)$ is $r \in (r_{\rm c},1)$ where $r_{\rm c}=r_{\rm c}(\beta)$ is a critical value given by
\begin{equation}
  r_{\rm c}(\beta) = -\frac{1}{K'_\beta(0)} = \frac{\zeta(\beta)}{\zeta(\beta-1)} .
  \label{rc}
\end{equation}
The saddle point solution (\ref{pr1},\ref{pr2}) 
holds for $r \in (r_{\rm c},1)$,
while for $r \in (0,r_{\rm c}]$ the solution is linear in $r$
\begin{equation}
  \phi(\beta,r) = r K_\beta(0) = r \log \zeta(\beta) .
  \label{phi_cond}
\end{equation}
For $r \in (r_{\rm c},1)$ the system is in the fluid phase
while for $r \in (0,r_{\rm c})$ in the condensed phase. 
In the condensed phase one box captures 
a finite fraction of all particles $S$ as 
$S\rightarrow \infty$ \cite{bbj}. This is a real-space condensation.
It should be noted that in the condensed phase the R\'enyi entropy is determined by the bulk part of the distribution, which remains in the critical state, since the contribution from the condensate in a single box vanishes in the thermodynamic limit (\ref{dlimit}). 
The phase transition, which the system undergoes for the given inverse temperature $\beta$ at the critical inverse density 
$r_{\rm c}(\beta)$ manifests as a singularity
of the thermodynamic potential $\phi(\beta,r)$. The singularity can 
be seen as a discontinuity of a derivative of the thermodynamic potential
\begin{equation}
\lim_{r \rightarrow r^+_{\rm c}} \partial^{n}_r \phi(\beta,r) \ne 
\lim_{r \rightarrow r^-_{\rm c}} \partial^{n}_r \phi(\beta,r).
\label{dd}
\end{equation}
Generically the discontinuity is infinite, as a result of the derivative divergence, but there are cases in which the discontinuity is finite - usually for the first order phase
transitions, but not only.
The transition is said to be $n$-th order when the $n$-th derivative $\partial^{n}_r \phi$ is discontinuous  while all $k$-th derivatives 
$\partial^{k}_r \phi$ for $k=1,\ldots, n-1$ are continuous at the 
critical point $r_{\rm c}(\beta)$.

\begin{figure}
    \centering
    \includegraphics[width=0.46\textwidth]{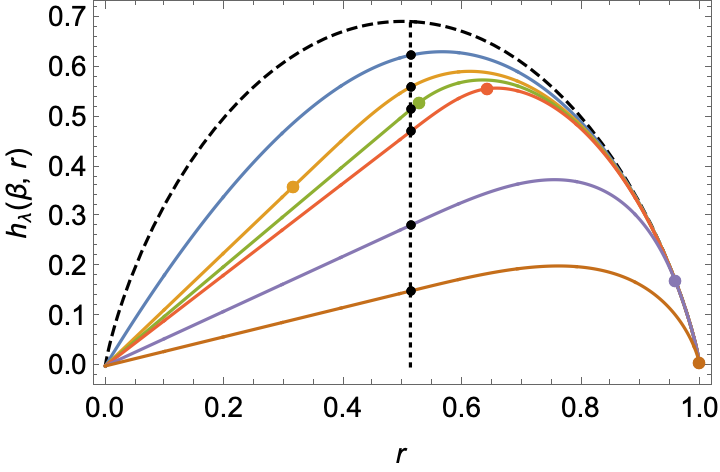} \qquad
    \includegraphics[width=0.46\textwidth]{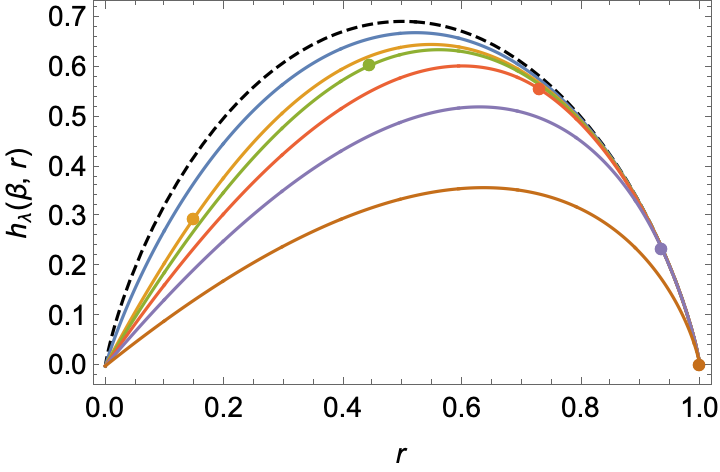}
    \caption{The left graph shows R\'{e}nyi entropy density $h_\lambda(\beta,r)$ for $\beta=5/2$
    and from top to bottom for $\lambda=0$ (black dashed), $0.6$ (blue), 
    $0.9$ (orange), $1.01$ (green), $1.1$ (red), $2.0$ 
    (purple) and $\infty$ (brown). The primary singular
    points lie on the dotted vertical line at 
    $r_{\rm c} = \zeta(5/2)/\zeta(3/2) \approx 0.5135$.
    They are marked by small black dots.
    The position of the secondary singular points depends
    on $\lambda$  (see Eq. \eqref{rcl}). The secondary singular points are marked by colored dots.
    For $\lambda=1.01$ the primary and secondary singular points almost merge. For $\lambda=0.6$
    there is no secondary singular point, because $\lambda \beta \le 2.0$. 
     Note that the curves are linear in the intervals $(0,r_*)$, where $r_*=\min(r_{\rm c},r_{\rm c,\lambda})$ \eqref{phi_cond}.  
    The right graph shows $h_\lambda(\beta,r)$ for $\beta=3/2$ and from top to bottom for $\lambda=0$ (black dashed), 
    $0.8$ (blue), $1.4$ (orange), $1.6$ (green), $2.0$ (red). $3.0$ (purple) and $\infty$ (brown). There are no primary singular points in this case. The secondary
    singular points, however, do exist for $\lambda>4/3$ and are marked by dots. There is no secondary singular point on the blue curve ($\lambda=0.8$), because $\lambda < 4/3$. The singular behaviour of the R\'{e}nyi entropy density $h_{\lambda}(\beta,r)$ is not visible by the naked eye in the figure as it is associated with the discontinuity (or divergence) of 
    higher derivatives $\partial_r^n h_{\lambda}(\beta,r)$ but not of the function itself.
    For instance, for $\lambda=3.0$ (purple), the singular point is at $r_{\rm c,\lambda} =\zeta(4.5)/\zeta(3.5) \approx 0.936$ \eqref{rcl}. At this point, the second derivative $\partial^2_r h(\beta,r)$ is discontinuous and the fourth derivative $\partial^4_r h(\beta,r) \sim (r-r_{c,\lambda})^{-1/2}$ is divergent, as follows from Eqs. \eqref{b3g} and \eqref{b3d}.}
    \label{h_figure}
\end{figure}

The parametric equations (\ref{pr1}, \ref{pr2}) can be used to plot 
the function $\phi(\beta,r)$ and thus also $h_\lambda(\beta,r)$ (\ref{hphi}). We show two examples in Fig. \ref{h_figure}
for $\beta=5/2$ and $\beta=3/2$ to illustrate the behaviour
for the two cases mentioned above. 
Here we are interested in singular points 
where the Rényi entropy density is singular, or more precisely
where any $n$-th derivative is discontinuous 
\begin{equation}
\lim_{r \rightarrow r^+_{\rm c}} \partial^{n}_r h_\lambda(\beta,r) \ne 
\lim_{r \rightarrow r^-_{\rm c}} \partial^{n}_r h_\lambda(\beta,r)
\end{equation}

The R\'enyi entropy density $h_\lambda(\beta,r)$ inherits its singularities from $\phi(\beta,r)$. The primary singularity
{\rm  lies at the critical point}: $r_{\rm c}(\beta)$ (\ref{rc}), but
$h_\lambda(\beta,r)$ may also have a secondary singularity 
coming from $\phi(\lambda \beta,r)$ in (\ref{hphi}) which is located 
at  a different point:
\begin{equation}
    r_{{\rm c},\lambda}(\beta) = \frac{\zeta(\lambda \beta)}{
    \zeta(\lambda \beta-1)} .
    \label{rcl}
\end{equation}
In this respect, the R\'enyi entropy density for the zeta-urn model is behaving (unsurprisingly, given (\ref{Esig},\ref{Zsig},\ref{ps})) as an equilibrium statistical mechanical system, with singularities at two different $\beta$ values. It was found in \cite{wbe} that this was not the case for the totally asymmetric exclusion process (TASEP), where $H_2$ was calculated by combinatorial means and found to possess no secondary singularities. It was suggested there that secondary singularities would generically be absent in such non-equilibrium systems, since they were a consequence of the relations (\ref{HZ},\ref{HF}), which are peculiar to equilibrium systems. Although the distribution of particles in a zeta-urn model can be considered as arising as a non-equilibrium steady state in a zero range process (ZRP) with suitable jumping rates for the particles \cite{zrp}, we are treating it as a purely equilibrium model here.

\begin{figure}
    \centering
    \includegraphics[width=0.44\textwidth]{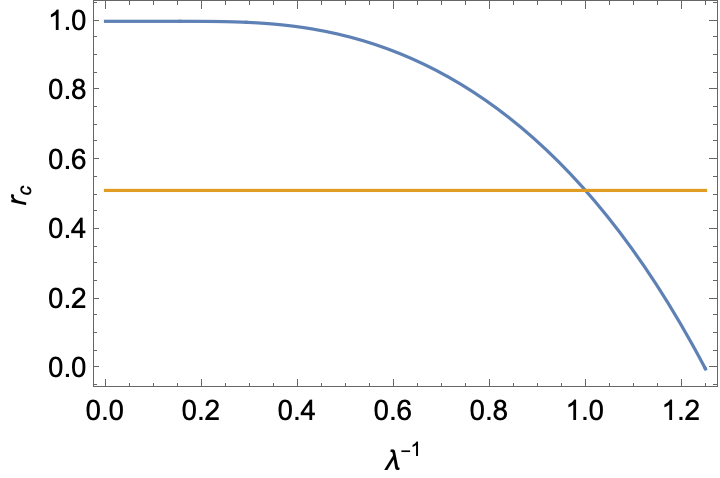} \qquad
    \includegraphics[width=0.44\textwidth]{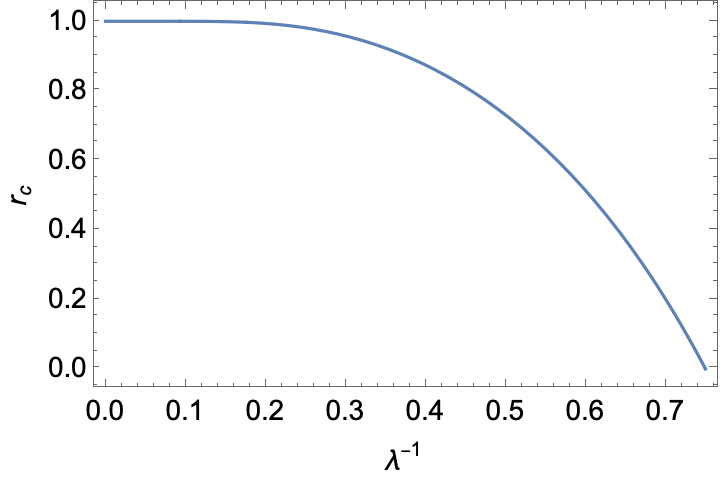}
    \caption{The positions of the primary (orange) and secondary (blue) singular points,  (Eqs. (\ref{rc}) and (\ref{rcl}), respectively) for $\beta=5/2$ (left) 
    and $\beta=3/2$ (right) plotted as a function of $\lambda^{-1}$
    in the range $\lambda^{-1} \in (0,\beta/2)$. 
    The blue curves
    have  the same universal shape, differing mainly in their
    support $(0,\beta/2)$, which depends on $\beta$.
    There are no secondary singular points outside this range. Similarly, there are no primary singular points for $\beta\le 2$.
    }
    \label{rc_figure}
\end{figure}

The function $h_\lambda(\beta,r)$ has a primary singularity at $r_{\rm c}$ (\ref{rc}) 
for $\beta>2$ and a secondary singularity at $r_{{\rm c},\lambda}$
(\ref{rcl}) for $\lambda \beta>2$, so there are four different 
cases: 
\begin{itemize}
\item[(a)] $h_\lambda(\beta,r)$ is regular for any $r\in (0,1)$ for
$\beta\le 2$ and $\lambda\beta \le 2$
\item[(b)] $h_\lambda(\beta,r)$ is
singular at $r_{\rm c,\lambda}$ for $\beta\le 2$ and
$\lambda\beta > 2$
\item[(c)] $h_\lambda(\beta,r)$ is
singular at $r_{\rm c}$ for $\beta > 2$ and $\lambda\beta \le 2$
\item[(d)] $h_\lambda(\beta,r)$ is
singular at $r_{\rm c}$ and $r_{{\rm c},\lambda}$ for
$\beta> 2$ and $\lambda\beta > 2$.
\end{itemize}
The positions of the secondary and primary singularities 
merge for $\lambda \rightarrow 1$ (the Shannon entropy). The behaviour is
illustrated in Fig. \ref{rc_figure}. 
The primary singularities of $h_\lambda(\beta,r)$ at $r_{\rm c}$ (\ref{rc}) are directly related to the singularities of 
the thermodynamic potential $\phi(\beta,r)$ at the critical point
$r=r_{\rm c}$ while the secondary singularities of $h_\lambda(\beta,r)$
\eqref{hphi} are related to  the singularities of $\phi(\lambda \beta,r)$ at
the phantom critical point $r_{\rm c,\lambda}$ \eqref{rcl}.
The thermodynamic potential $\phi(\lambda \beta,r)$ 
has a critical point $r_{\rm c}(\beta)$ for $\beta>2$, 
and $\phi(\lambda \beta,r)$ has a phantom critical point $r_{\rm c,\lambda}(\beta)$ for $\lambda \beta > 2$.

The critical behaviour of $\phi(\beta,r)$ is encoded in
discontinuities of higher order derivatives of $\phi(\beta,r)$ 
at the critical point $r_{\rm c}$. The second derivative 
 for $r\rightarrow r_{\rm c}^+$  behaves like (see Appendix)
\begin{equation}
    \partial^2_r \phi(\beta, r) \sim \left\{ 
    \begin{array}{lcl} -c_1 (r - r_{\rm c})^x + \ldots & {\rm for} & \beta \in (2,3) \\
    +c_2 \log(r - r_{\rm c}) + \ldots
    & {\rm for} & \beta =3 \\
    - c_3 + \ldots & {\rm for} & \beta \in (3,\infty) \end{array}
    \right.  
\end{equation}
where $x=(3-\beta)/(\beta-2)$ (\ref{x}),
and $c_1,c_2,c_3$ are positive constants. Dots indicate sub-leading
terms. On the other hand, $\partial^2_r \phi(\beta, r)=0$ for 
$r\rightarrow r_{\rm c}^-$. So we conclude, that 
the second derivative has a finite discontinuity for $\beta \in (3,\infty)$ and it is logarithmically divergent for $\beta=3$. It is continuous for $\beta \in (2,3)$ but then higher derivatives diverge.
Moreover, as discussed in the Appendix, for $\beta \in (3,\infty)$,
the second derivative contains, among the sub-leading terms, 
a term $\sim (r-r_{\rm c})^{\beta-3}$ for a non-integer $\beta$ or 
a term $\sim (r-r_{\rm c})^{\beta-3} \log(r-r_{\rm c})$ for an integer $\beta$. This term leads to a divergence of
higher derivatives for $r\rightarrow r_{\rm c}^+$.

To summarize, the second derivative of the R\'{e}nyi entropy density (\ref{hphi}) inherits its singular behaviour at the critical point from
$\partial_r^2 \phi$ 
\begin{equation}
    \partial_{r}^2 h_\lambda(\beta,r) = 
    \frac{\partial_{r}^2 \phi(\lambda \beta,r) - \lambda
    \partial_{r}^2 \phi(\beta,r)}{1-\lambda} .
    \label{hphi2}
\end{equation}
This is the primary singularity.  However, 
additionally, $\partial_{r}^2 h_\lambda(\beta,r)$ can acquire 
a secondary singularity at $r_{{\rm c},\lambda}$ (\ref{rcl}) 
when $\lambda \beta>2$. The singularity type is the same as
for the primary singularity, except that it corresponds to the critical 
behavior of the thermodynamic potential $\phi$ for the inverse temperature  
$\lambda \beta$ rather than $\beta$.

For $\lambda=0$, $h_0(\beta,r)$ (\ref{h0}) is independent of $\beta$ and
has no singular points in the range $r\in (0,1)$. Another
exception is $\lambda \rightarrow 1$ because the resulting
singularity of $h_1(\beta,r)$ comes from the merging of
the primary and the secondary singularities. One can expect
from Eq. (\ref{h1}) that the power-law singularities
will acquire an extra logarithmic factor from 
the derivative of a power depending on $\beta$.
For example, the second derivative has the following singularity  
\begin{equation}
    \partial^2_{r} h_1(\beta,r) \sim 
    (r-r_{\rm c})^{x} \log(r-r_{\rm c}) .
\end{equation}
The logarithm here is generated from the derivative in the second term in (\ref{h1}).

\section{Discussion}

We have calculated analytically
the R\'enyi entropy for the abundance vector
defined by the Gibbs measure for a zeta-urn model. In the thermodynamic
limit for a suitable choice of parameters the model has a (condensation) phase transition, which
manifests as a singularity of the free energy at the critical point.
The R\'enyi entropy also has a singularity at this point, but in addition to
this  the R\'enyi entropy can, depending on its order, display a secondary singularity at another point  unlike the archetypal non-equilibrium model, the TASEP, as demonstrated in \cite{wbe}. 
The secondary singularity is a phantom of the 
original singularity but itself is not directly related to any critical behaviour in the system. The mechanism which leads
to the occurrence of the secondary singularity is quite generic, following from (\ref{HZ},\ref{HF}),
so such secondary singularities will occur
in other statistical mechanical models with Gibbs weights and phase transitions.  In the case of the TASEP the weights are given by matrix products and do not have this structure.
We stress, however, that these secondary singularities are rooted in the mathematical definition 
of the R\'enyi entropy rather than in physical behaviour of the 
system. Below we illustrate this by a discussion of the secondary
singularities for the R\'enyi divergence, where they
arise from a comparison of {\em two} systems at different temperatures, 
so as such they cannot be a physical property of either one of the systems individually.

The R\'enyi divergence of order $\lambda$ 
\begin{equation}
    \Delta_\lambda(\bar{p}|\bar{q}) = 
    \frac{1}{\lambda-1}\log \sum_\sigma \frac{p_\sigma^\lambda}{q_\sigma^{\lambda-1}}  
\end{equation}
is a generalisation of the Kullback-Leibler divergence \cite{kl}
(which is reproduced from the expression above in the limit $\lambda \rightarrow 1$). The R\'enyi divergence for two Gibbs distributions 
$\bar{p} =\{p_\sigma = e^{-\beta_1 E_\sigma}/Z(\beta_1), 
\sigma=1,\ldots, \mathcal{N}\}$ and
$\bar{q} =\{q_\sigma = e^{-\beta_2 E_\sigma}/Z(\beta_2), 
\sigma=1,\ldots, \mathcal{N}\}$, for the same statistical system
at different temperatures $T_1=1/\beta_1$ and $T_2=1/\beta_2$ is
\begin{equation}
    \Delta_\lambda(\beta_1|\beta_2) = \frac{1}{\lambda-1}\log
    \frac{Z(\lambda\beta_1 - (\lambda-1) \beta_2)Z^{\lambda-1}(\beta_2)}{Z^\lambda(\beta_1)} 
    \label{delta}
\end{equation}
which should be compared with (\ref{HZ}).
For convenience, we have replaced arguments of $\Delta_\lambda$ 
by $\beta_1$ and $\beta_2$ which uniquely identify the thermal
distributions $\bar{p}$ and $\bar{q}$. 
The divergence is proportional to the temperature difference
and the heat capacity of the system
\begin{equation}
    \Delta_\lambda(\beta|\beta+\Delta \beta) = 
    \frac{\lambda \Delta \beta^2}{2} \left( \log 
    Z(\beta)\right)'' + o(\Delta \beta^2) .
\end{equation}
For the zeta-urn model, in the thermodynamic limit (\ref{dlimit}),
Eq. (\ref{delta}) yields 
\begin{align}
    \Delta_\lambda(\beta_1|\beta_2) = \frac{1}{\lambda-1}
    \biggl\{\phi\left(\lambda\beta_1 - (\lambda-1) \beta_2,r\right) -
    \lambda \phi (\beta_1,r)  
     + (\lambda-1)\phi(\beta_2,r)\biggr\} .
\end{align}
We see that apart from the primary singularities at 
$r_{{\rm c},1}=\zeta(\beta_1)/\zeta(\beta_1-1)$ and 
$r_{{\rm c},2}=\zeta(\beta_2)/\zeta(\beta_2-1)$,
the R\'enyi divergence can have a secondary singularity at
$r_{{\rm c},\lambda}=\zeta(\beta_\lambda)/\zeta(\beta_\lambda-1)$, where 
$\beta_\lambda = \lambda \beta_1 - (\lambda-1)\beta_2$,  
if $\beta_\lambda>2$.

Returning to the interpretation of the R\'enyi entropy density as a diversity measure, it is interesting to look at the behaviour of $h_{\lambda}$ as $r$ is varied in Fig. \ref{h_figure}. For a given $\beta$ as $r$ is {\em decreased} (i.e. as the density of particles is increased) $h_{\lambda}$ initially increases, reaching a maximum value at
\begin{equation}
    \partial_{r} h_\lambda(\beta,r) = 
    \frac{\partial_{r} \phi(\lambda \beta,r) - \lambda
    \partial_{r} \phi(\beta,r)}{1-\lambda}    = 0 
    \label{hphi1}
\end{equation}
with the limiting case of $\lambda=1$ being given by
\begin{equation}
    \partial_{r} h_1(\beta,r) = \partial_r \phi(\beta,r)-\partial_{\beta r}^2 \phi(\beta,r)  = 0    \, .
    \label{hphi11}
\end{equation}
The values taken by $\partial_r \phi(\beta,r)$ 
in (\ref{hphi1},\ref{hphi11}) will depend on whether a singularity has been encountered or not, giving $\log \zeta(\beta)$
for $r \in (0,r_{\rm c}]$ and
$K_\beta(\alpha(r))$ for $r \in (r_{\rm c},1)$, with similar considerations for $\partial_r \phi(\lambda \beta,r)$. 
Whether the maximum value of $h_{\lambda} (\beta,r)$ is attained as $r$ is decreased before a singularity is encountered will depend on both $\lambda$ and $\beta$. For instance, when $\beta=5/2$ we can see in Fig. \ref{h_figure} that the maximum of $h_{\lambda}$ occurs before any singularities are encountered when $\lambda \le1$, whereas it may encounter the secondary singularity at $r_{{\rm c},\lambda}(\beta)$ for sufficiently large $\lambda$ before reaching its maximum. The Shannon entropy, $h_1$, attains a maximum value in the fluid phase and then decreases linearly with $r$ into the condensed phase after encountering the primary singularity  as the density of particles is increased. In the second example in   Fig. \ref{h_figure}, $\beta=3/2$, there is no primary singularity but the secondary one exists for $\lambda>4/3$ and can lie on either side of the maximum of $h_{\lambda}$ depending on the value of $\lambda$.

It is also clear that whatever the value of $\beta$, the maximum value of $h_{\lambda}$ decreases from that of $h_0$ as $\lambda$ is increased and shifts to larger $r$. The value  of the ``maximum diversity'' and the density at which it occurs hence both depend on the value of $\lambda$ chosen for a given $\beta$. Similarly, increasing $\beta$ for a given $\lambda$ decreases the maximum value of $h_{\lambda}$ and shifts it to larger $r$. The task of maximizing the diversity for a zeta-urn model in the ensemble we consider thus depends both on what we mean by the diversity, e.g the choice of $\lambda$, and what parameters we have under our control, e.g. $r$ and/or $\beta$.

\appendix

\section{} \label{app}

In the Appendix, we discuss the critical behavior of the thermodynamic potential
$\phi(\beta,r)$ at $r=r_{\rm c}$. We want to establish how the singularity type 
at the critical point $r_{\rm c}$ depends on $\beta$. We find it convenient to take the partial derivative of $\phi(\beta,r)$ with respect to $r$ because the corresponding 
parametric equations for $\partial_r \phi(\beta,r)$ are simpler than those for $\phi(\beta,r)$ (\ref{pr1},\ref{pr2}) and are therefore more useful in the analysis of critical point singularity. For $r \in (0,r_{\rm c}]$ we get 
\begin{equation}
  \partial_r \phi(\beta,r) = K_\beta(0) = \log \zeta(\beta).
\end{equation}
while for $r \in (r_{\rm c},1)$
\begin{equation}
    r = -\frac{1}{K_\beta'(\alpha)}
    \label{prp1}
\end{equation}
and
\begin{equation} 
    \partial_r \phi(\beta,r) = K_\beta(\alpha)
    \label{prp2}
\end{equation}
where $\alpha \in (0,\infty)$. The equations will be used as follows.
First we will expand the right hand side of (\ref{prp1}) to
extract the dependence of $\alpha =\alpha(r)$ on $r$, for $r$ 
approaching $r_{\rm c}$ from above. Then we will
substitute $\alpha=\alpha(r)$ into the expression on the right hand side of
(\ref{prp2}) to determine the type of singularity of $\partial_r \phi(\beta,r)$ for $r\rightarrow r_{\rm c}$. To this end we will use the series expansion of the polylogarithm for a non-integer $\beta$ \cite{as}: 
\begin{equation}
   {\rm Li}_\beta(e^{-\alpha}) = 
   \Gamma(1-\beta) \alpha^{\beta-1} + \sum_{k=0}^\infty
   \frac{(-1)^k \zeta(\beta-k)}{k!} \alpha^k .
   \label{exp-non-int}
\end{equation}
For $\beta \in (2,3)$, (\ref{prp1}) and (\ref{prp2}) can be written as
\begin{equation}
    r = r_{\rm c} + B \alpha^{\beta-2} + o\left(\alpha^{\beta-2}\right)
\end{equation}
and
\begin{equation} 
    \partial_r \phi(\beta,r) = \mu_{\rm c} -a_1 \alpha + o\left(\alpha\right) 
\end{equation}
with coefficients $r_{\rm c}$, $\mu_{\rm c}$, $a_1$ 
and $B>0$, depending on $\beta$. The dependence of the coefficients
on $\beta$ can be easily determined (for instance 
$\mu_{\rm c}(\beta)=\log \zeta(\beta)$), but will not be displayed in the analysis below, because we want to concentrate on the dependence 
on $r$. From the first equation we get
\begin{equation}
    \alpha = C (r-r_{\rm c})^{\frac{1}{\beta-2}} + o\left((r-r_{\rm c})^{\frac{1}{\beta-2}}\right)
\end{equation}
with $C=B^{-1/(\beta-2)}$. Substituting this into the second equation 
leads to
\begin{equation} 
    \partial_r \phi(\beta,r) = \mu_{\rm c} - D (r-r_{\rm c})^{\frac{1}{\beta-2}} + 
     o\left((r-r_{\rm c})^{\frac{1}{\beta-2}}\right)
\end{equation}
with $D=a_1 C$. The coefficients $\mu_{\rm c}$ and $D$ depend only 
on $\beta$, so if we take the
derivative of both sides with respect to $r$ we get
\begin{equation} 
    \partial^2_{r} \phi(\beta,r) = -E (r-r_{\rm c})^{\frac{3-\beta}{\beta-2}} + 
     o\left((r-r_{\rm c})^{\frac{3-\beta}{\beta-2}}\right) 
\end{equation}
with $E=(\beta-2)D$ being a positive function of $\beta \in (2,3)$. 
The second derivative $\partial^2_{r} \phi(\beta,r)$ is related to
particle density fluctuations. For $\beta \in (2,3)$ the exponent 
\begin{equation}
    x= \frac{3-\beta}{\beta-2}
    \label{x}
\end{equation} 
changes from zero to infinity when $\beta$ changes from $3$ to $2$, so the transition is of second or higher order. For $\beta=2$ the transition disappears and there is no phase transition for $\beta\le 2$. On the other hand, for $\beta=3$ the second derivative has a logarithmic singularity
at $r_{\rm c}$. To see this, let us use the series expansion of the polylogarithm for an integer $\beta$ \cite{as} 
\begin{eqnarray}
   {\rm Li}_\beta(e^{-\alpha}) = 
   \frac{(-1)^{\beta-1}}{(\beta-1)!}\left(H_{\beta-1}-\log(\alpha)\right) \alpha^{\beta-1} 
   +
    \sum_{k=0,k\ne \tau-1}^\infty
   \frac{(-1)^k \zeta(\beta-k)}{k!} \alpha^k 
   \label{exp-int}
\end{eqnarray}
with $H_n=1+1/2+\ldots+1/n$ being the $n$-th harmonic number. 
For $\beta=3$, Eqs. (\ref{prp1}) and (\ref{prp2}) take the form
\begin{equation}
    r = r_{\rm c} + b_1 \alpha + B\alpha \log \alpha + 
    o\left(\alpha \log \alpha \right)
\end{equation}
and
\begin{equation}
\partial_r \phi(\beta,r) = \mu_{\rm c} - a_1 \alpha + 
o\left(\alpha\right)
\end{equation}
where again the coefficients $r_{\rm c}$, $b_1$, $B$, 
$\mu_{\rm c}$ and $a_1$ depend only on $\beta$. Calculating
$\alpha$ as a function of $r$ from the first equation we get
\begin{eqnarray}
    \alpha &=& c_1(r - r_{\rm c}) - C (r-r_{\rm c}) 
    \log (r-r_{\rm c}) \nonumber \\
    &+&
    o\left((r-r_{\rm c}) 
    \log (r-r_{\rm c})\right)
\end{eqnarray}
with $c_1 = 1/b_1 + B/b_1^2 \log b_1$ and $C=B/b_1^2$.
Substituting this into the second equation we get
\begin{eqnarray}
    \partial_r \phi(\beta,r) = \mu_{\rm c} - d_1 (r-r_{\rm c}) 
    + D (r-r_{\rm c}) \log (r-r_{\rm c}) 
    + 
    o\left((r-r_{\rm c}) \log (r-r_{\rm c})\right) .
\end{eqnarray}
where $d_1=a_1 c_1$ and $D=a_1 C$. As a consequence
the second derivative has a logarithmic singularity for $r\rightarrow r_{\rm c}^+$
\begin{equation}
    \partial_{r}^2 \phi(\beta,r) = -d_1 + D + D \log (r-r_{\rm c}) + 
    o\left(\log (r-r_{\rm c})\right) . 
    \label{logs}
\end{equation}
What is essential in the last equation is that the second derivative 
diverges logarithmically when $r \rightarrow r_{\rm c}^+$. This means that
for $\beta=3$ the particle density fluctuations are infinite
at the critical point: $r\rightarrow r_{\rm c}^+$. 

For $\beta>3$, Eqs. (\ref{prp1}) and (\ref{prp2}) take the form
\begin{equation}
    r = r_{\rm c} + b_1 \alpha + \ldots + B \alpha^{\beta-2} + 
    o(\alpha^{\beta-2})
\end{equation}
and
\begin{equation}
\partial_r \phi(\beta,r) = \mu_{\rm c} -  a_1 \alpha + o\left(\alpha\right),
\end{equation}
so for $r\rightarrow r_{\rm c}^+$
\begin{eqnarray}
\partial_r \phi(\beta,r) = \mu_{\rm c} - \frac{a_1}{b_1} (r-r_{\rm c}) + 
\ldots + \frac{a_1 B}{b_1^{\beta-1}} (r-r_{\rm c})^{\beta-2} 
+ 
o\left((r-r_{\rm c})^{\beta-2}\right) 
\label{b3g}
\end{eqnarray}
and hence 
\begin{equation}
\lim_{r\rightarrow r_{\rm c}^+} \partial^2_{r} \phi(\beta,r) = 
-\frac{a_1}{b_1}<0 .
\label{b3d}
\end{equation}
On the other hand 
\begin{equation}
\lim_{r\rightarrow r_{\rm c}^-} \partial^2_{r} \phi(\beta,r) = 0
\end{equation}
as follows from (\ref{phi_cond}). Hence the second derivative
has a finite discontinuity for $\beta>3$. Additionally, we see that
the first derivative contains a singular 
term $\sim (r-r_{\rm c})^{\beta-2}$, and therefore the second derivative
contains a term $\sim (r-r_{\rm c})^{\beta-3}$ which makes higher derivatives diverge for $r\rightarrow r_{\rm c}^+$. 
For an integer $\beta$ this term is $\sim (r-r_{\rm c})^{\beta-3} \log(r-r_{\rm c})$, as follows from (\ref{exp-int}).

\end{document}